# Репутационная система для онлайн-сообществ

Антон Колонин[1,2], Бен Герцель[2], Дебора Дуонг[2], Мэтт Айкл[2], Нэйт Знидар[2]

[1]Aigents, Новосибирск, Россия
[2]ФондStitching SingularityNET, Амстердам, Нидерланды
{anton, ben, deborah, matt}@singularitynet.io

**Аннотация.** Понимание принципов консенсуса в сообществах и нахождение путей для нахождения решений оптимальных сообщества в целом становится критически важным по мере увеличения скоростей и масштабов взаимодействия в современных распределенных системах. Такими системами могут являться как социально-информационные компьютерные сети, объединяющие массы людей, так и много-агентные вычислительные платформы, в том-числе — одно-ранговые, действующие на основе распределенных реестров. Наконец, в настоящее время становится возможным появление гибридных экосистем, которые включают и человека, и компьютерные системы с использованием искусственного интеллекта. Мы предлагаем новую форму консенсуса для подобных систем, основанную на репутации участников, вычисляемой согласно принципу «текучей демократии». Полагаем, что такая система будет иметь большей сопротивляемостью к социальной инженерии и манипуляцией репутацией, чем у действующих систем. В этой статье мы обсуждаем основные принципы и варианты реализации такой системы, а также представляем предварительные практические результаты.

**Ключевые слова**: коллективный интеллект, консенсус, одно-ранговые вычисления, распределенные системы, репутация, социальные вычисления, текучая демократия

## 1   Введение и предпосылки

Начиная с появления децентрализованных и распределенных компьютерных систем без централизованного управления, стало понятно, что надежность определения репутации участников представляет собой серьезную проблему, и эта проблема была объектом всестороннего изучения [1]. Надежное решение по определению репутации узла вычислительной сети оказывается критичным для одно-ранговых систем, в которых каждый узел может взаимодействовать с любым другим узлом в сети [2]. Стандартные теоретические основы для нахождения решения связаны с так называемой «задачей византийских генералов», где переменное число участников с переменными уровнями доверия независимо голосует для достижения консенсуса по решению, которое вносится в публичный реестр, чтобы все сообщество узнало об этом решении и чтобы это решение было для него благоприятным [3]. Поскольку уровень доверия каждого узла системы заранее неизвестен, необходимо снизить риск воздействия на принятие решения со стороны узлов-«предателей», которые пытаются повлиять на консенсус в пользу враждебной части сообщества, одержав верх над остальными членами. В существующих распределенных вычислительных системах, основанных на технологии блокчейн, применяются различные алгоритмы достижения консенсуса, использующие различные формы взвешенного голосования, каждая из которых предлагает определенную эвристику, какое качество узла в компьютерной сети можно использовать, чтобы догадаться о его ожидаемом уровне доверия [4].

Для построения любых систем коллективного интеллекта или просто коллективного принятия решения при большом числе участников и отсутствию жесткой иерархической управляющей структуры («вертикали власти»), необходимы как достаточная автономность в выработке собственных вариантов решений каждым и участников, так и возможность быстрого и надежного определения общественного консенсуса в пределах всей системы. Для предотвращения злоупотреблений и манипуляций в подобной распределенной много-агентной сети необходима высококачественная репутационная система. Например, в создаваемой системе общего искусственного интеллекта (ОИИ) SingularityNET, обеспечение высоко-надежного репутации требует искусственного интеллекта (ИИ) само по себе, что ведет к взаимной рекурсии между ОИИ и оценкой репутации в распределенных системах ИИ. Таким образом, умение построить надежную систему исчисления репутации участников является критически важным для решения проблем ОИИ, и любая парадигма ОИИ должна тем или иным образом решать "проблему доверия" в отношениях между элементами много-агентной системы. Мы полагаем, что решение состоит из двух частей - А) относительно простой базовый алгоритм для определения репутации как «уровня доверия» между участниками в простых общих случаях и Б) система, которая занимается верификацией оценок на основе базового алгоритма на основе методов ИИ, для более сложных случаев. Из описания очевидно, что подобное решение могло бы быть востребовано не только для обеспечений консенсуса в системах распределенного ИИ, но и в существующих человеческих онлайн-сообществах, а также в грядущих смешанных человеко-машинных экосистемах.

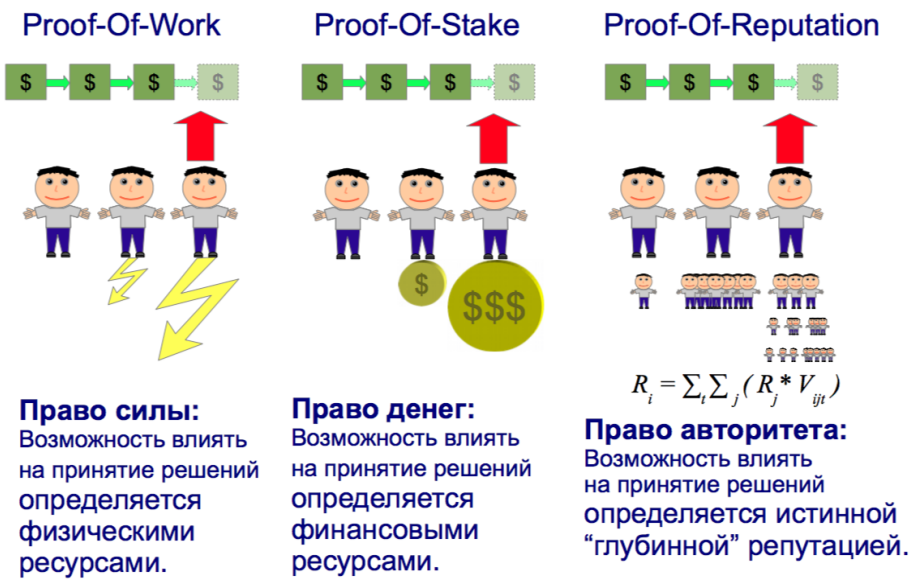

**Рис.1.** Типы консенсуса в распределенных системах .

Большинство алгоритмов достижения консенсуса, обсуждавшихся в предыдущих работах [4] и внедренных в существующие популярные распределенные вычислительные системы, такие как Ethereum и Bitcoin, подвержены взлому при известных обстоятельствах. Алгоритмом достижения консенсуса, называемым «доказательство права работой» или Proof-of-Work (POW), где участник системы голосует вычислительной мощностью, которой он владеет, может злоупотребить общество союзников, которые смогут в нужный момент сконцентрировать свыше 51% компьютерной мощности для достижения консенсуса в свою пользу. С исторической точки зрения на человеческое сообщество, данный алгоритм соответствует «праву силы», характерному для большинства древних сообществ и животных стай и стад. Еще один известный алгоритм достижения консенсуса, называемый «доказательство права финансовой ставкой» или «Proof-of-Stake» (POS), подразумевает голосование суммой финансовых средств, которыми владеет каждый участник. Это аналогично консенсусу в современных капиталистических обществах с «правом

денег», где «богатый всегда богатеет». Такое решение приводит постепенно к глубокому разрыву в доходах, при котором со временем один участник или группа участников, сконцентрировавших достаточные средства, могут повлиять на достижение консенсуса исключительно в личных интересах, а не интересах всего сообщества. Усовершенствованная версия POS, называемая «делегирование права доказательства финансовой ставкой» или «Delegated-Proof-of-Stake» (DPOS), предполагается как решение последней проблемы явным делегированием права управлять «делегатам», назначенным участниками, владеющими большими долями, но это лишь приводит к ручному контролю над распределенной системой.

В настоящей работе мы предлагаем, по нашему мнению, более перспективную версию алгоритма достижения консенсуса, называемую «доказательство права репутацией» или «Proof-of-Reputation» (POR), которая может быть использована для построения «репутационного консенсуса» или консенсуса, основанного на репутации. В этом случае, возможности участника повлиять на достижение консенсуса могут определяться суммой репутации, социальным капиталом или «кармой», фактически заработанной участником в процессе взаимодействия с другими участниками в указанный период времени, принимая во внимание репутации самих этих участников [5]. Описываемый ниже алгоритм реализует это на основе принципа «текучей демократии», где голоса одних участников неявно влияют на голоса других, перетекая в них, и называется алгоритмом «текучего ранжирования» или «liquid rank».

Побочным эффектом возможности вычислить надежный социальный консенсус в сообществе участников является способность выявить надежность каждого из участников, поддерживая наиболее эффективные и безопасные коммуникации между ними. Мы указывали на это в нашей предыдущей работе по разработке системы репутации для децентрализованного рынка услуг искусственного интеллекта, называемого SingularityNET [6]. Доказательство права репутацией предоставляет возможность измерить и отследить динамику развития репутации каждого участника общества. Это может быть применимо к любому обществу интеллектуальных агентов, реальных людей, взаимодействующих в социально-информационных сетях в среде Интернету, или даже гибридным человеко-машинным сообществам [7]. Вычислительные решения построенные на основе описываемой технологии могут быть отнесены к классу «социальных вычислений» либо сама социальная среда рассматривается как средство распределенных вычислений репутаций каждого элемента этой среды.

## 2 Алгоритм расчета репутации

Описанный ниже алгоритм вычисления репутации основан на методах вычисления количественных характеристик элементов графов в социальных сетях, опубликованных в более ранних работах [7,8,9], адаптированных к проблеме расчета социальных репутаций для выстраивания лучшего алгоритма достижения консенсуса [3,4,5,6].

Репутацию $R_i(t)$ участника общества $i$ в момент времени $t$ можно вычислить инкрементально, на основании его репутации в предшествующий момент времени $R_i(t-1)$ либо, в случае отсутствия таковой, определенной репутации по умолчанию $Rd$, принятой за его изначальную репутацию. Изменения в репутации участника $i$ могут быть вызваны различного рода рейтингами (оценками), выпущенными множеством других участников $j$ в отношении конкретного аспекта репутаций $k$ и конкретной предметной категории репутации $c$. Предполагается, что аспект $k$ это стандартный критерий типа скорости обслуживания, качества или доброжелательности, в то время как категория $c$ может указывать на область компетенции участника, например, живопись, финансовый прогноз или доставка пиццы.

Рейтинги можно разделить на два вида. Первый – удостоверяющие рейтинги типа «ставка» $S_{ijkc}$, которые могут присутствовать или отсутствовать в любой момент времени $t$, которые $j$ может присуждать $i$ или отбирать у него. Далее, существуют транзакционные рейтинги $F_{ijkce}$, которые можно регистрировать в истории взаимодействий участников и которые ассоциируются либо с финансовыми транзакциями от $j$ к $i$ (финансовые рейтинги), либо с актами голосования (рейтинги голосования) в отношении конкретных событий $e(t)$, таких как публикации, посты, комментарии, номинации или задачи и обязанности, которые имеет $i$ по отношению к $j$. Большинство рейтингов могут быть явными или неявными. Явные рейтинги голосования имеют оценочное значение, выраженное как позитивное, негативное или в виде любого числа на определенной шкале, в то время как неявные представляют собой комментарии и обзоры $j$ в отношении $i$, где истинное значение рейтинга нужно определять из среды, используемой для комментария или обзора, такой как текст на естественном языке. Удостоверяющие рейтинги $S_{ijkc}$ могут подкрепляться стоимостью финансовой доли $Q_{ij}$ или нет. В последнем случае удостоверяющие рейтинги соответствуют «подписке» или «дружбе» в социальных сетях. Рейтинги голосования по транзакциям $F_{ijkce}$ могут подкрепляться финансовой ценностью $G_{ije}$ или нет. Например, ценность голоса участника $j$ в отношении квалификации поставщика услуг $i$ может усиливаться с учетом стоимости всей услуги $e(t)$ в рамках системы, где учитываются сделки по услугам или поставкам товаров и вычисляются репутации поставщиков. Если же голосование не подкреплено финансовой ценностью, то оно соответствует отметке «нравится» в социальны сетях.

Значения рейтинга можно определять в границах от *-1* до *1* для негативных и позитивных рейтингов, а для целей визуализации их можно определять от *-5* до *5*, от *0* до *10* или в любых других границах, которые кажутся удобными для восприятия в той или иной системе. Что касается финансовых рейтингов, то проведение упоминаемых ниже опытов с блокчейном Ethereum показало, что желательно стандартизировать нелинейные распределения финансовой ценности транзакций с помощью логарифма следующим образом:

$F'_{ijce} = log_{10}(F'_{ijce})/MAX(log_{10}(F'_{ijce}))$

Рейтинги для различных аспектов $k$ можно смешивать, чтобы сделать вывод об общей репутации, используя системный параметр смешивания $H_k$. Затем следующие формулы помогут выявить дифференциальную репутацию в момент времени $t_n$ как относительное улучшение репутации из-за удостоверяющие $dS_i(t_{n-1},t_n)$ и транзакционного $dF_i(t_{n-1},t_n)$ компонентов, где $t$ для событий $e(t)$ варьируется в пределах от $t_{n-1}$ до $t_n$.

$dS_i(t_{n-1},t_n) = \Sigma_k(H_k*\Sigma_{jct}(S_{ijkc}(t_n)*Q_{ijc}(t_n)*R_j(t_{n-1}))) / \Sigma_{jct}(Q_{ijc}(t_n)*R_j(t_{n-1})))/\Sigma_k(H_k)$

$dF_i(t_{n-1},t_n) = \Sigma_k(H_k*\Sigma_{jct}(F_{ijkce}(t)*G_{ijce}(t)*R_j(t_{n-1}))) / \Sigma_{jct}(G_{ijce}(t)*R_j(t_{n-1})))/\Sigma_k(H_k)$

В упрощенном виде, когда не рассматриваются никакие аспекты или категории, улучшение репутации одобрения или транзакции можно упростить, как указано ниже.

$dS_i(t_{n-1},t_n) = \Sigma_{jt}(S_{ij}(t_n)*Q_{ij}(t_n)*R_j(t_{n-1})) / \Sigma_{jt}(Q_{ij}(t_n)*R_j(t_{n-1}))$

$dF_i(t_{n-1},t_n) = \Sigma_{jt}(F_{ije}(t)*G_{ije}(t)*R_j(t_{n-1})) / \Sigma_{jct}(G_{ije}(t)*R_j(t_{n-1}))$

На практике можно применять либо удостоверяющую репутацию, либо транзакционную репутацию. Если используются обе, смешанное улучшение репутации можно рассчитать, используя смешивающие коэффициенты $S$ и $F$ для каждого вида репутаций, соответственно.

$dP_i(t_{n-1},t_n) = (S * dS_i(t_{n-1},t_n) + F * dF_i(t_{n-1},t_n)) / (S + F)$

Дифференциальную репутацию можно дополнительно нормализовать, максимально улучшив лучшую репутацию за период времени в интервале от 0.0 до 1.0:

$P_i(t_{n-1},t_n) = dP_i(t_{n-1},t_n) / MAX_i(ABS(dP_i(t_{n-1},t_n)))$

Основываясь на репутации, заработанной в предыдущий период с $t_o$ до $t_{n-1}$, новую репутацию за последнее время $t_n$ можно рассчитать, смешав предыдущую ценность репутации с дифференциальной.

$R_i(t_n) = ((t_{n-1} - t_o) * R_i(t_{n-1}) + (t_n - t_{n-1}) * P_i(t_{n-1},t_n)) / (t_n - t_o)$

Как выяснилось во время экспериментов, которые мы обсудим ниже, линейное вычисление репутации при его применении к экспериментальным сообществам дает достаточно нелинейное распределение ценностей репутации в сообществе, когда высокие оценки имеют очень немногие члены, а у остальных членов сообщества репутации равны нулю. Для улучшения распределения для практических целей неотрицательную логарифмическую дифференциальную репутацию можно рассчитать следующим образом, поэтому в двух вышеуказанных формулах $lP_i(t_{n-1},t_n)$ можно использовать вместо $dP_i(t_{n-1},t_n)$.

$lP_i(t_{n-1},t_n) = SIGN(dP_i(t_{n-1},t_n)) * log_{10}(1+ABS(dP_i(t_{n-1},t_n)))$

Нашу методику оценки репутации можно преобразовать, чтобы заработанная репутация ухудшалась быстрее или медленнее. Мы можем применить дополнительные смешивающие коэффициенты к самому последнему важному интервалу времени и к более ранним интервалам времени при расчете $R_i(t_n)$, чтобы ценность ранее заработанной репутации могла ухудшаться быстрее или медленнее после того, как её скорректировали с последней дифференциальной репутацией.

Также возможно рассчитать более детальную структуру репутации, для различных аспектов или категорий, как мы покажем ниже на примере транзакционной дифференциальной репутации. На основании этих идей можно рассчитать более точные репутации $R_{ic}(t_n)$, $R_{ik}(t_n)$ и $R_{ikc}(t_n)$ внутри общества.

$dF_{ic}(t_{n-1},t_n) = \Sigma_k(H_k * \Sigma_{jt}(F_{ijkce}(t) * G_{ijce}(t) * R_j(t_{n-1}))) / \Sigma_{jt}(G_{ijce}(t) * R_j(t_{n-1})))/\Sigma_k(H_k)$

$dF_{ik}(t_{n-1},t_n) = \Sigma_{jct}(F_{ijkce}(t) * G_{ijce}(t) * R_j(t_{n-1})) / \Sigma_{jct}(G_{ijce}(t) * R_j(t_{n-1}))$

$dF_{ikc}(t_{n-1},t_n) = \Sigma_{jt}(F_{ijkce}(t) * G_{ijce}(t) * R_j(t_{n-1})) / \Sigma_{jt}(G_{ijce}(t) * R_j(t_{n-1}))$

## 3    Варианты реализации и практического применения

Положения и элементы алгоритма, предложенные выше, могут быть реализованы и использованы многими способами, в зависимости от решений, принятых относительно временных интервалов расчета репутации, а также вариантов обеспечения вычислений и хранения полученных данных, как будет рассмотрено ниже. В конце мы введем понятия «**Репутационный консенсус**» и «**Доказательство права репутацией**» и «**Репутационный майнинг**».

*Определение временных интервалов*
Работа репутационной системы будет зависеть от определения границ времени на основании охвата времени между циклами оценки репутации во временной промежуток от $t_{n-1}$ до $t_n$.

С одной стороны, возможен перерасчет «**за весь период существования**», когда учитываются все рейтинги между $t_0$ и $t_n$. В этом случае есть возможность учесть прошлые изменения в истории рейтинга при последующих перерасчетах. Однако это гораздо дороже и занимает больше времени. Также в этом случае нельзя достичь ухудшения репутации, как это было указано выше, и потребуется усложнение функций дифференциальной репутации, что приведет к появлению дополнительной, ограниченной во времени весовой функции, что будет придавать больший вес более недавним рейтингам.

С другой стороны, может быть «**мгновенно-инкрементальный**» перерасчет, где временные интервалы между $t_0$ и $t_n$ соответствуют интервалам между последующими транзакциями, поэтому каждая транзакция влияет на глобальное изменение репутации. В этом случае отсутствует отсрочка в изменении репутации, но применение этого в распределенной среде может оказаться совсем не тривиальным. Вместе с тем это может оказаться благоприятным для распределенных систем, не основанных на технологии блокчейн.

Может быть реализован и «**практически-инкрементальный**» перерасчет, промежуточный между предыдущими двумя, где временные интервалы между $t_0$ и $t_n$ – это года, кварталы, месяцы, недели, дни и т.д. Это более эффективный и быстрый способ, однако, изменение репутации может быть отсрочено, устареть ближе к концу интервала перерасчета.

И, наконец, есть нечто среднее между двумя последними способами, такое как «**поблочно-инкрементальный**» перерасчет, при котором используются блоки последних последующих транзакций для определения временного интервала. Внедрение этого способа в распределенные системы на основе блокчейн может оказаться вполне практичным.

*Варианты обеспечения репутационных вычислений*

С точки зрения поддержания существуют централизованный, децентрализованный и распределенный варианты.

При «**централизованном**» варианте все репутации рассчитываются одной стороной или системой, которую назначило и которой доверяет сообщество, называемое Централизованное Репутационное Агентство.

При «**децентрализованном**» варианте все репутации рассчитываются многими сторонами или системами, которым эту функцию доверило сообщество, причем система из нескольких участвующих сторон может называться Децентрализованных Репутационным Агентством. В этом случае существуют еще два варианта. Первый - когда все агентства должны достичь консенсуса по текущему состоянию репутации, поэтому их можно назвать Децентрализованным Координированным Репутационным Агентством. Второй вариант – когда все они независимо поддерживают репутации, и в этом случае они будут являться Независимыми Децентрализованными Репутационными Агентствами.

В случае «**распределенного**» варианта все участники сообщества отвечают за расчет репутаций, например, как в блокчейне, где все участники сообщества могут отвечать за регистрацию и проверку транзакций. В этом случае также возможно либо достичь скоординированного консенсуса всех участников по состоянию репутаций («координированное агенство»), либо каждый будет иметь личную точку зрения на репутацию других («независимые агенства»).

*Варианты хранения репутационных данных*

При любом из вышеописанных вариантов возможно несколько подходов к хранению репутационных данных, а именно, они могут быть транзитными, с локальным постоянного хранением или глобальным постоянным хранением децентрализованным или распределенным способом.

В «**транзитном**» варианте все репутации всегда рассчитываются «на лету» вместе с доступными данными по удостоверениям и транзакциям. В случае «**локального постоянного хранения**» все вычисляемые репутации хранятся в локальном хранилище специально сформированного Репутационного Агентства или любого участника системы, без необходимости синхронизировать данные по статусу репутационной системы с другими участниками. Что касается «**глобального постоянного хранения**», то репутационные данные в этом случае поддерживаются децентрализованным или распределенным способом во всех Репутационных Агентствах или всеми участниками системы.

*Децентрализованный репутационный консенсус*

Если использовать разработки из нашей более ранней работы [6], небезынтересно рассмотреть, как различные Репутационные Агентства достигают децентрализованного консенсуса в отношении совместно используемых репутационных данных. Предлагается следующий алгоритм администрирования этого процесса.

1. Во время каждого цикла расчета репутации, каждое агенство отправляет другим свою версию статуса репутационной системы или информацию о состоянии репутационных данных, определяя репутацию для всех известных участников.

2. Для всех получателей в течение цикла каждое последующее полученное состояние после первого должно быть идентичным предыдущему.

3. Если очередное получаемое состояние не равно предыдущему, тогда набор последующих состояний отмечается как оспариваемый и в администрацию/службу текущего контроля отправляется предупреждение.

4. Как только получен требуемый системный минимум послдедовательных идентичных состояний, состояние отмечается как достоверное, и больше состояния не принимаются.

5. Когда получен указанный системный максимум не идентичных состояний и имеется минимальное количество идентичных состояний, тогда состояние, которое поддерживается большинством отправлений, отмечается как достоверное, больше состояния не принимаются, и в администрацию/службу текущего контроля отправляется предупреждение, идентифицирующее отправителей не идентичного состояния.

6. Если в течение установленного системного периода, начиная с первого отправления, не происходит описанное в п. 4 и п. 5, консенсус считается нарушенным, никакие репутации не обновляются, а в администрацию/службу текущего контроля отправляется предупреждение, что нужно проверить всю репутационную систему.

7. В случае применения вышеуказанного процесса в распределенных системах, применяющих алгоритм достижения репутационного консенсуса для «доказательства права репутацией» или «Proof-of-Reputation» (POR) согласно нижеприведённому определению, голосование за принятое состояние репутации можно проводить с учетом репутации самих Агентств репутации с соответствующими изменениями в урегулировании спора, как это указано в п. 5 выше.

*Доказательство права репутацией*

В случае, когда сообщество, которое применяет расчеты репутации, проводит распределенные вычисления, необходимые для распределенного консенсуса в рамках системы на основе распределенного реестра, такой как блокчейн, вычисленная репутация может использоваться для целей достижения консенсуса в самой системе. Уровень репутации, заработанной конкретным узлом блокчейна, может влиять на его возможность повлиять на формирование блока таким же образом, как имеющаяся вычислительная мощность в случае Proof-of-Work или «сумма финансовой ставки» в случае Proof-of-Stake. Такого рода консенсус можно назвать «доказательство права репутацией» или «Proof-of-Reputation» (POR).

*Майнинг репутации*

Если репутации поддерживаются при распределенной реализации, когда их вычисляют посредством участия каждого участника сообщества, тогда это можно считать процессом «майнинга». Чтобы это произошло, состояния репутации, отправленные участниками, должны соответствовать протоколу децентрализованного репутационного консенсуса, описанного выше. В этом случае участник или несколько первых участников, которые первыми отправили соответствующие состояния репутации, признанные остальными участниками, имеют право получить компенсацию от сообщества, как это происходит в алгоритме консенсуса Proof-of-Work посредством решения криптографических задач, не имеющих никакой практической ценности.

## 4 Апробация в социальных сетях и блокчейнах

В процессе изучения возможностей практических вычислений репутации в существующих социальных сетях и блокчейнах, проводились эксперименты на платформе социальных вычислений Aigents [7,9] с использованием социальных сетей, таких как Facebook и Google+, а также социальных сетей на основе технологии блокчейн, таких как Steemit и Ethereum. Ниже мы подведем некоторые итоги нашей работы.

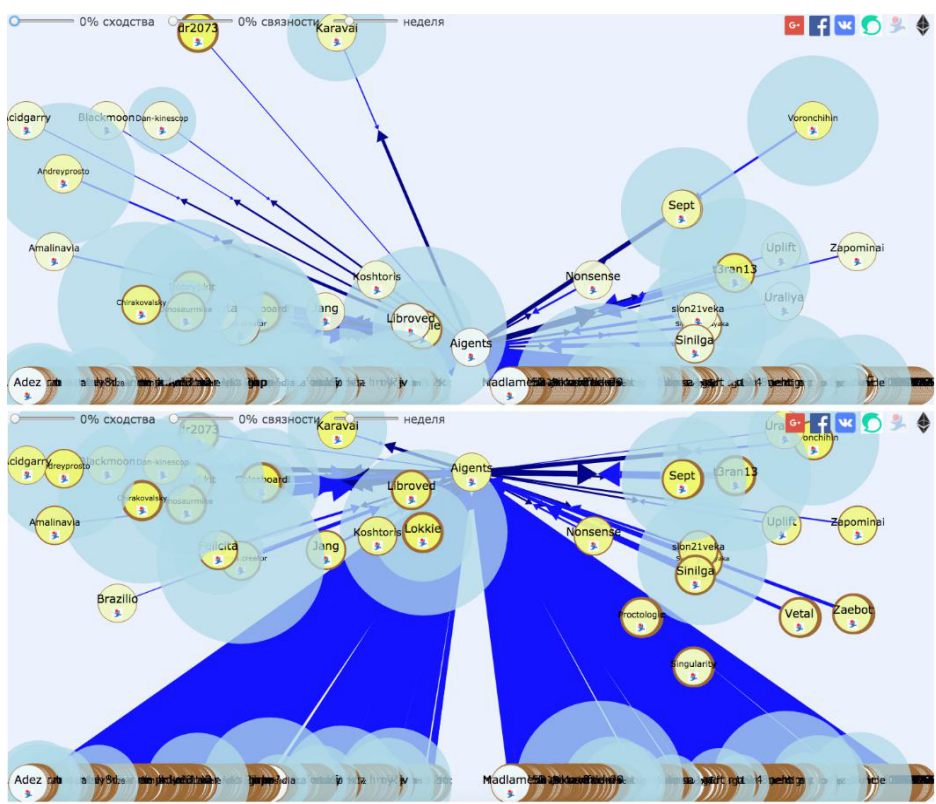

**Рис. 2.** Распределение участников сообщества в социальной сети на блокчейн Steemit для ближайшего окружения выбранного «центрального» пользователя за выбранный интервал времени период на основании явных и неявных рейтингов. Вертикальная позиция узла графа соответствует уровню репутации участника в данном ограниченном сообществе; размер кружка-«ореола» вокруг узла представляет сходство между соответствующим участником и участником в центре графа; толщина стрелок между узлами соответствует интенсивности явных и неявных рейтинговых взаимодействий между участниками. Верхняя диаграмма построена без логарифмической нормализации, нижняя с логарифмированием рассчитанных репутаций.

*Facebook и Google+*
Эти социальные сети, используемые платформой Aigents, позволяют оценивать репутации ближайшего социального окружения и пользователей систем, основанных на взаимодействии пользователя и других равноправных пользователей на своей домашней странице. Результаты исследований ограниченного круга пользователей, которые согласились поделиться своим опытом, показывают значительную надежность в случае достаточно интенсивного общения. С другой стороны, если большая часть общения пользователя происходит в группах или на домашних страницах других пользователей, официальные программные интерфейсы социальных сетей ограничивают доступ к этим данным, в связи с чем надежность результатов падает. Также, к сожалению исследователей и разработчиков приложений, в настоящее время Facebook практически полностью ограничил официальный доступ к данным пользователей для разработчиков приложений, так что в настоящее время возможности подобных исследований за пределами самой сети Facebook практически равны нулю. Из-за ограничений конфиденциальности, результаты анализа данных в этой статье приведены быть не могут.

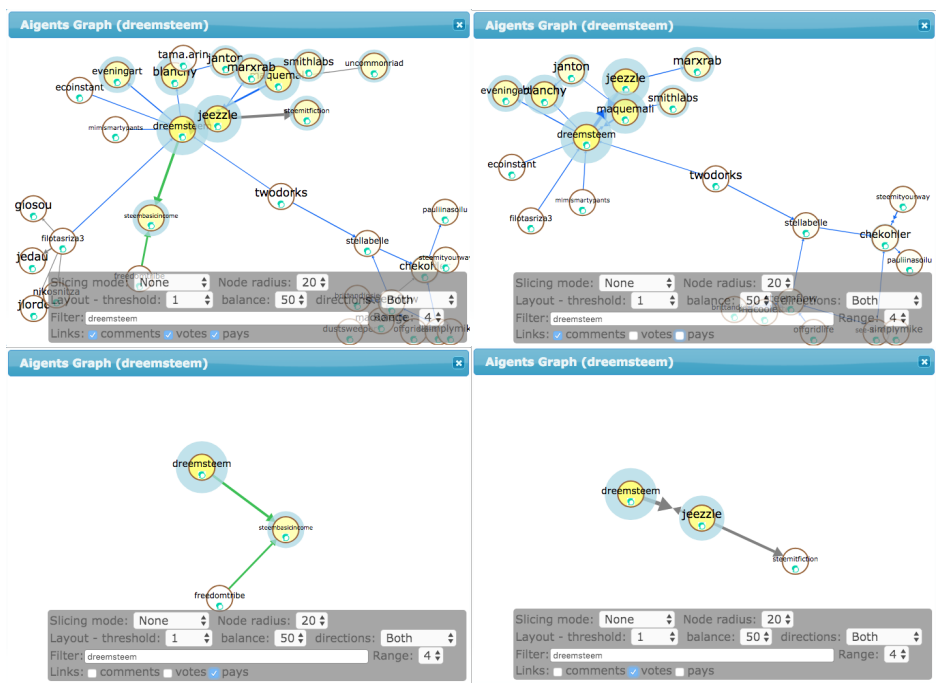

**Рис. 3.** Сетевой график отдельного сообщества Steemit вокруг нескольких «лидеров мнений» - на расстоянии до четырех сетевых переходов («рукопожатий»). В левом верхнему углу — использованы все типы рейтингов; в правом верхнем — только комментарии; в левом нижнем — только финансовые транзакции; в правом нижнем — только голосования.

*Steemit*
Поскольку социальная сеть Steemit основана на технологии публичного блокчейна, все её результаты находятся в открытом доступе и представляют возможность для исследования. Было проведено изучение многочисленных публичных аккаунтов на предмет вычислений репутаций по сетевым социальным графам, полученным из блокчейна. В этом случае явные рейтинги транзакций, такие как голосование за сообщения и комментарии, а также подразумеваемые рейтинги, такие как непрямые комментарии, использовались в упрощенном форме. То есть финансовая ценность в Steemit не ассоциировалась с рейтингами, и подразумеваемая ценность комментария считалась положительной вне зависимости от эмоционального тона комментария. Изучение показало, что такая упрощенная форма хорошо соответствует ожидаемой реальности, в соответствии с рассмотрением истории взаимодействия реальных участников. Попутно обнаружилось, что использование репутаций с логарифмической зависимостью позволяет более четко идентифицировать социальную структуру, как это показано на Рис. 2.

Другое исследование на данных Steemit проводилось с использованием методов сетевого анализа с выборочным включением тех или иных видов рейтингов — включая финансовые транзакции, голосования и комментарии в различных комбинациях — результаты исследования представлены на Рис 3. В отличии от Рис. 2, относительная репутация участника сети в сообществе определялась не вертикальным положением, а только насыщенностью цвета, а графическое пространство позволяет отражать структуру взаимодействия между участниками сообщества и его внутреннюю иерархию.

*Ethereum*

В блокчейне Ethereum доступны только финансовые рейтинги транзакций, и мы использовали их для вычисления изменения репутаций участников сети на всем протяжении жизненного цикла Ethereum. Было обнаружено, что распределение ценностей финансовой транзакции слишком нелинейно и без применения операции логарифмирования в формуле расчета репутации распределение участников происходит по схеме «победитель забирает все», то есть несколько лидеров оказываются в самом верху репутационной иерархии, а все остальные оказываются примерно на одном уровне около нуля. Поэтому нам пришлось подключить логарифмическое масштабирование финансовых транзакций как на уровне стоимости каждой транзакции, так и на уровне вычисления репутационных состояний по каждому интервалу времени. В результате оказалось возможно получить правдоподобные, хотя и существенно нелинейные распределения репутации участников.

Для оценки достоверности полученных результатов составлялся «эталонный список» участников с «ожидаемыми» значениями репутации, где адреса потенциально «опасных» счетов (по данным https://etherscamdb.info/scams/) помечались значением 0.0, а адреса потенциально «благонадежных» участников (по данным https://etherscan.io/) помечались значением 1.0. Репутации из данного списка сопоставлялись с расчетными репутациями участников экосистемы Ethereum расчетом коэффициента корреляции по Пирсону. На сегодняшний день, результаты можно считать очень предварительными, так как на сегодняшний день не все варианты параметров были опробованы. Тем не менее, различные варианты дают стабильно положительные значения коэффициента корреляции между указанными списками, что указывает на возможность автоматического расчета ожидаемой репутации участников экосистемы в реальном времени.

## 5   Выводы и заключение

Данная статья является переводом нашей более ранней работы [10] с небольшими дополнениями в части результатов, полученных за прошедшие полгода, а также нижеследующих выводов.

- Представлена исчерпывающая модель вычисления репутаций в много-агентных системах, основанная на различного рода исторических данных, представляющих разнообразные способы взаимодействия между агентами.
- Предложены различные способы построения репутационных систем, в зависимости от конкретных условий и требований.
- Часть предложенных разработок доступна в виде вычислительной платформы персональной социальной аналитики Aigents на вебсайте https://aigents.com а также в открытом коде платформы Aigents: https://github.com/aigents/aigents-java.
- Развитие репутационной системы и исследование её возможностей для распределенных экосистем и корпоративных систем ведется в рамках проекта по созданию экосистемы сервисов искусственного интеллекта с открытым кодом SingularityNET: https://github.com/singnet/reputation.
- Продолжение работ идет как как в области имитационного моделирования, для изучения устойчивости репутационной системы репутации и репутационного консенсуса к различного рода манипуляциям и видам мошенничества направленным на умышленное искажение репутаций участников в пользу атакующих.
- Ожидается дальнейшее развитие распределенных репутационных систем и репутационных агентов, основанных на обсуждаемых принципах, а также их дальнейшее усовершенствование существующих.

## Благодарности

## Ссылки